\documentclass[twocolumn,groupedaddress,superscriptaddress,amsmath,amssymb,floatfix,aps,prl,showpacs]{revtex4}
\usepackage{amsmath}    
\usepackage{graphicx}   
\usepackage{verbatim}   
\usepackage{color}      
\usepackage{subfigure}  
\usepackage{hyperref}   
\usepackage{soul}
\usepackage{hyphenat}

\begin{document}
\bibliographystyle{prsty} 

\title{Fundamental Limits on Sensing Chemical Concentrations with Linear Biochemical Networks}
\author{Christopher C. Govern}
\author{Pieter Rein ten Wolde}
\affiliation{FOM Institute AMOLF, Science Park 104, 1098 XG Amsterdam, The Netherlands}

\begin{abstract}
  Living cells often need to extract information from biochemical
  signals that are noisy. We study how accurately cells can
    measure chemical concentrations with signaling networks that are
  linear. For stationary signals of long duration, they can reach, but
  not beat, the Berg-Purcell limit, which relies on uniformly
  averaging in time the fluctuations in the input signal. For short
  times or nonstationary signals, however, they can beat the
  Berg-Purcell limit, by non-uniformly time-averaging the
  input. We derive the optimal weighting function for time
    averaging and use it to provide the fundamental limit of measuring
    chemical concentrations with linear signaling networks.
\end{abstract}


\maketitle
Cells measure concentrations of chemicals via receptors on their
surface. These measurements, however, are inevitably corrupted by
noise that arises from the stochastic arrival of ligand molecules at
the receptor by diffusion and from the stochastic binding of the
ligand to the receptor. Biochemical networks that transmit the
information on the ligand concentration from the surface of the cell
to its interior often have to filter this noise extrinsic to the cell as
much as possible. However, how the capacity of signaling networks to
remove extrinsic noise depends on their design, and what the
fundamental limits to this capacity are, remain open questions.

Several studies have addressed the question how accurately 
 the ligand
concentration $c$ can be estimated from the time trace of the number of
ligand-bound receptors, $S(t)$, over some integration time
$T$ \cite{berg1977,bialek2005,levinepre2007,
  levineprl2008,wingreen2009,endres2010,levineprl2010,mora2010}. Berg and Purcell assumed that
the estimate $\hat{c}$ with least error is the one that matches
the observed time average of the stochastic signal $S(t)$, $\bar{S} =
1/T \int_0^T S(t) dt$, giving all the signal values equal weight
in the average \cite{berg1977}.  When $S(t)$ is stationary, with mean $\mu_S$,
variance $\sigma^2_{S}$, and correlations that decay exponentially
over a time $\tau_{\rm c}$, the estimate $\hat{c}$ has variance (error)
\cite{levinepre2007,postma2007, wingreen2011}:
\begin{eqnarray}
\label{eq:BP}
\sigma_{\hat{c}}^2 [ \bar{S} ] &=& \frac{\sigma_{\bar{S}}^2}{ \left( d\mu_{S}/dc \right)^2} 
= \frac{\sigma^2_{S} / \left( d\mu_S / dc \right)^2}{\frac{T}{2 \tau_{\rm c}} \left(1-\frac{1-\exp(-T/\tau_{\rm c})}{T/\tau_{\rm c}}\right)^{-1}}.
\end{eqnarray}
More recently, Mora, Endres, and Wingreen showed that, when $T \gg \tau_c$, maximum likelihood estimation produces an estimate that is better by 50\%, since the
time-average includes noise from stochastic ligand unbinding, which
provides no information about the ligand concentration
\cite{wingreen2009,mora2010}.

While these previous studies have considered how much information
about the ligand concentration is encoded in receptor-occupancy time
traces, they do not address the question how much information
biochemical networks can actually extract from these time traces. To
extract all the information, the biochemical networks downstream of
the receptors would need to construct a maximum likelihood estimate
\cite{wingreen2009,mora2010}. However, it is not clear that typical
biochemical networks do this, nor is it clear that they time-average
signals {\em uniformly} as in the Berg-Purcell estimate.  Moreover, the
previous analyses \cite{berg1977,bialek2005,levinepre2007,
  levineprl2008,wingreen2009,endres2010,levineprl2010,mora2010}
assumed an integration time $T$, but what time scales in the
processing network actually set the integration time remains unclear.
We therefore study how accurately biochemical networks can estimate
the ligand concentration from receptor time traces.

We focus on a simple but broad class of signaling networks, linear
networks \cite{rapoport}.  Many networks respond
linearly over the range of fluctuations in their input
(e.g. \cite{wiggins2007,sorin2006,deronde2010}) and a systematic study can be done analytically.  Since the
effects of noise intrinsic to the molecular interactions inside cells
have been well studied \cite{tostevin2010,
  tkacik2008,deronde2010,paulsson2003}, we focus on networks in the
deterministic limit. This enables us to understand the unique effects due to the noise in
the input signal.

Linear networks time-average the input signal, but do not generally
give rise to uniformly weighted time-averages.  We
study how different signaling motifs sculpt the weighting of the
signal as a function of time, and how this affects the precision of
ligand sensing. While linear networks cannot extract all of the
information in the input signal (i.e. the maximum likelihood estimate
\cite{wingreen2009}), they can, surprisingly, reach the Berg-Purcell
limit and even exceed it by 12\%; this
is because the optimal weighting is non-uniform, in contrast to the Berg-Purcell estimate. We show that a
simple network based on a feed-forward loop, a common motif in
biochemical networks \cite{shen-orr02}, can reach the bound for linear
signaling networks, and we elucidate the combination of time
  scales that sets the effective integration time.  We conclude by
studying how reliably biochemical signaling networks can extract
information from non-stationary signals.

We consider a cell that responds after a finite time $T_{\rm{o}}$ to a
change in its environment which happens at time $t=0$. This time
  $T_{\rm{o}}$ is the observation time, which, as we discuss below,
  provides an upper bound to the integration time.  As before, the
receptor time trace provides the signal to the cell, $S(t)$.  To compare
to previous results, we initially assume that the change in the
environment, and therefore the ligand concentration, is instantaneous,
and that the receptors immediately adjust. Moreover, we assume that
the fluctuations in $S(t)$ decay exponentially with correlation time
$\tau_{\rm c}$
\cite{levinepnas2008,derondethesis}. We neglect stochasticity in the
time $T_{\rm{o}}$, and, as mentioned above, the intrinsic noise in the
processing network. The capacity of the cell to respond is then
limited by the information in the stationary input $S(t)$ from
time $0$ to $T_{\rm{o}}$ (Fig. 1a).

As a measure of how much information the cell can extract, we
determine how accurately the ligand concentration can be estimated
from the molecular output $X$ of the processing network at the time
$T_{\rm{o}}$ of the response, assuming the response is made instantaneously
based on $X(T_{\rm{o}})$.  As illustrated below in examples, the output of a
linear signaling network is $X(T_{\rm{o}}) = \int_{-\infty}^{T_{\rm{o}}} f(T_{\rm{o}}-t^\prime)
S(t^\prime) dt^\prime$, where the unnormalized weighting (response or transfer) function
$f(\Delta t = T_{\rm{o}}-t^\prime)$ reflects how the processing network at time $T_{\rm{o}}$
weights the signal at an earlier time $t^\prime$ \cite{arkin2002}.  To compare to previous results, we
assume that either: (1) $f(T_{\rm{o}}-t)=0$ for $t<0$,
which corresponds to a scenario where the response time $\tau_{\rm
    r}$ of the network
is shorter than $T_{\rm{o}}$, or, equivalently, the network reaches steady state
by the time $T_{\rm{o}}$ (Fig. 1b); or (2) $S(t)=0$ for $t<0$, which
corresponds to a scenario in which the cell is initially in a basal
state (Fig. 1c).  In both cases, we then have $X(T_{\rm{o}}) = \int_0^{T_{\rm{o}}}
f(T_{\rm{o}}-t^\prime) S(t^\prime) dt^\prime$.  When neither $f(T_{\rm{o}}-t)$ nor
$S(t)$ are zero for $t<0$, then previous states of the environment,
corresponding to $t<0$, influence the state of the signaling network
at time $T_{\rm{o}}$.  Such previous environmental states can be a source of
additional noise in $X(T_{\rm{o}})$, complicating inference of the current
environmental state, as well as a source of information, helping
inference, if environmental transitions are correlated.

\begin{figure}
\includegraphics[width=8.5cm]{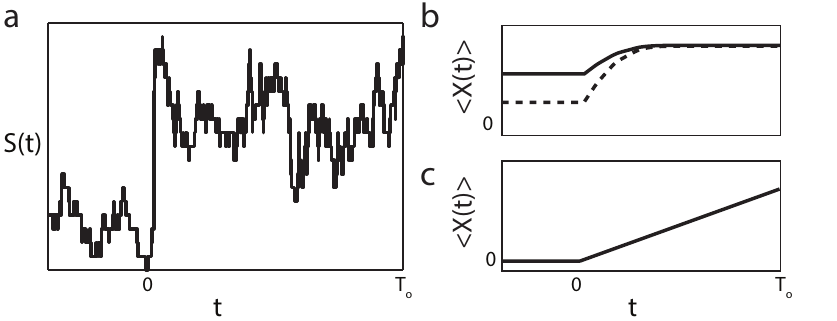}
\caption{ Responding to noisy environments.  (a) The environment
  changes instantaneously at time $t=0$, and the number of bound
  receptors, $S(t)$, adjusts instantaneously.  $S(t)$ is stationary between time 0 and the time $T_{\rm{o}}$, when the cell
  responds.  The signaling network is either in a steady state by
  time $T_{\rm{o}}$, independent of the initial condition} (b), or in a basal state at time $t=0$ (c). $X(t)$ denotes the number of $X$ molecules at time $t$.  The solid and dashed lines in panel b represent different initial conditions.
\label{fig:model2}
\end{figure}

\begin{figure*}
\includegraphics[width=17cm]{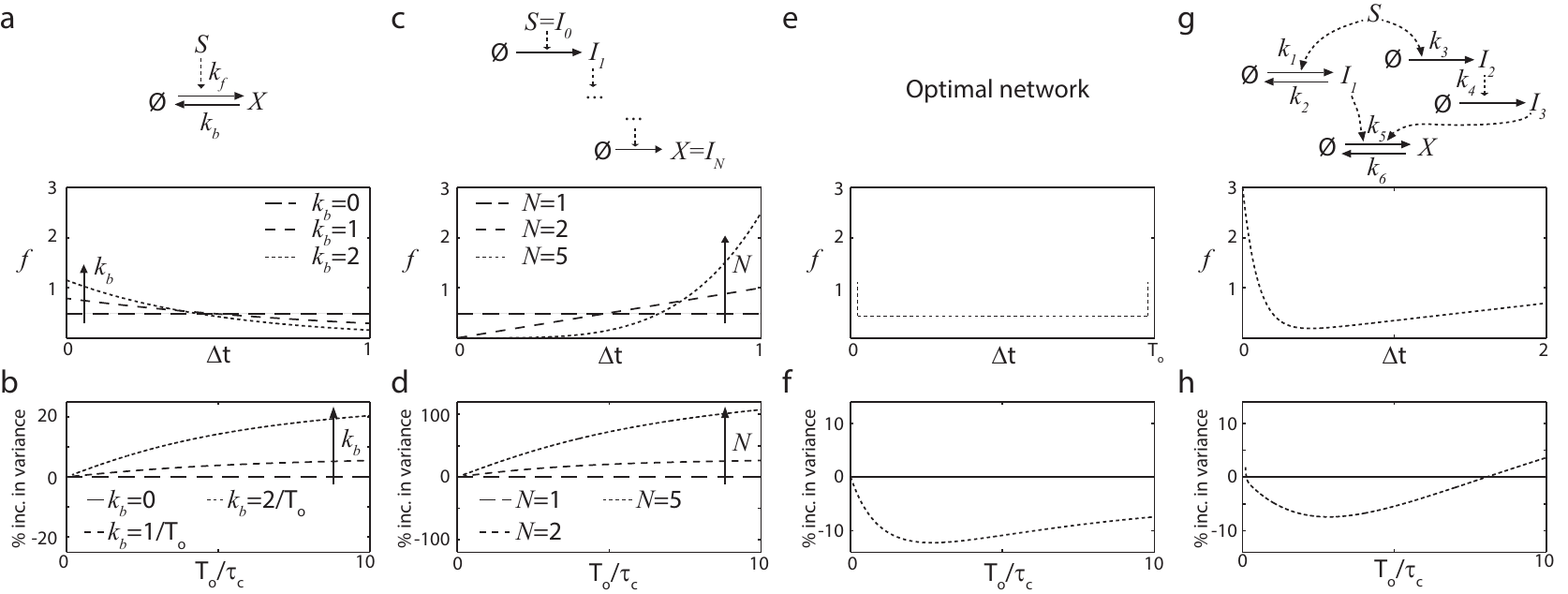}
\caption{ Extracting information from noisy input signals with linear
  signaling networks.  (a,c,e,g) The weighting functions corresponding
  to different signaling networks are not uniform. (b,d,f,h) The
    ability of a signaling network to measure ligand concentration
  depends on its weighting function.  The typical error (variance) in
  the estimate of ligand concentration is plotted as a percentage
  increase over the error of an estimate based on uniform weighting,
  assumed in the Berg-Purcell limit (Eq. 1 with $T=T_o$).  (a) Reversible, one-level
  cascades selectively amplify late ($t=T_{\rm{o}}$) values of the signal, (b)
  leading to worse performance than the uniform average.  (c)
  Irreversible, $N$-level cascades amplify early ($t=0$) values of the
  signal, (d) leading to worse performance than the uniform
  average. (e) The optimal weighting function, given by
  Eq. \ref{eq:f_opt}, averages the signal, selectively amplifying less
  correlated values.  The delta functions are truncated for
  illustration. (f) The optimal weighting function outperforms the
  uniform average. (g) A signaling network consisting of two branches,
  which selectively amplify late ($t=T_{\rm{o}}$) (left branch) and early
  ($t=0$) (right branch) values of the signal, approximates the
  optimal weighting function ($k_1 = 4.4 k_3 k_4 T_{\rm{o}}$; $k_2 = 20/T_{\rm{o}}$; $k_4 =
  0.35 k_3$; $\bar{f}$ independent of $k_3$,$k_5$; $k_6 \gg T_{\rm{o}}^{-1}$).
(h) The network in (f) can outperform the uniform average.  \label{fig:2} }
\label{fig:model3}
\end{figure*}

We start by considering a simple linear signaling network, a
reversible one-level cascade, in which the output molecule $X$ is
directly activated by the receptor with rate constant $k_{\rm f}$ and can be
degraded with rate constant $k_{\rm b}$ (Fig. \ref{fig:model3}a).  Then, deterministically,
$dX/dt=k_{\rm f} S - k_{\rm b} X$.  The response of this network at
time $T_{\rm{o}}$ is $X(T_{\rm{o}}) = \int_0^{T_{\rm{o}}} f(T_{\rm{o}}-t) S(t) dt + g(T_{\rm{o}}) X(0)$ with $
f(\Delta t) = k_{\rm f} \exp(-k_{\rm b} \Delta t)$ and $g(T_{\rm{o}}) = 
\exp(-k_{\rm b} T_{\rm{o}})$ (Fig. \ref{fig:model3}a).  We neglect the
term $g(T_{\rm{o}}) X(0)$ for the
reasons mentioned above: either because $T_{\rm{o}}$ is larger than the
response time $\tau_{\rm r} = 1/k_{\rm b}$ in which case $g(T_{\rm{o}}) \approx 0$, or,
because the initial state is ligand-free and $X(0) \approx 0$. We note that the weighting function $f(\Delta t)$ decays with increasing $\Delta t$, which
means that more weight is placed on more recent values of the input
signal. This is because the decay reaction is least likely to have
degraded the most recently produced $X$ molecules.

We now address the question how the departure from uniform weighting
affects the error in the estimate of the concentration. 
  Following the derivation of Eq. \ref{eq:BP}
\cite{berg1977}, an estimate of the ligand concentration from $X(T_{\rm{o}})$
has variance
\begin{equation}
\label{eq:sigma_c_1}
\sigma_{\hat{c}}^2 [X(T_{\rm{o}})] =  \sigma_{X(T_{\rm{o}})}^2 / \left(d\mu_{X(T_{\rm{o}})}/dc \right)^2,
\end{equation}
where the mean $\mu_{X(T_{\rm{o}})}$ of $X(T_{\rm{o}})$ is a linear function of $c$ over
the range of fluctuations in $X(T_{\rm{o}})$. Using $X(T_{\rm{o}}) =
\int_0^{T_{\rm{o}}} f(T_{\rm{o}}-t^\prime) S(t^\prime)dt^\prime$, we then arrive at (see supplement)
\begin{align}
\label{eq:sigma_c_2}
&\sigma_{\hat{c}}^2 [X(T_{\rm{o}})] =  \sigma_{\hat{c}}^2 [ S ]  \nonumber \\ &\times \int_0^{T_{\rm{o}}} \int_0^{T_{\rm{o}}} \bar{f}(T_{\rm{o}}-t_1) \bar{C}(t_1,t_2) \bar{f}(T_{\rm{o}}-t_2) dt_1 dt_2.
\end{align}
Here, $\sigma_{\hat{c}}^2 [ S ] = \sigma^2_{S} / \left( d\mu_S/dc
\right)^2$ is the error of an estimate based on an {\em instantaneous}
observation of the signal $S(t)$.  The reduction in error, resulting
from averaging the fluctuations in the input signal over time, depends on the normalized correlation function of the input fluctuations,
$\bar{C}(t_1,t_2)=\exp(-|t_2-t_1|/\tau_{\rm c})$, and on the
normalized weighting function, $\bar{f}(\Delta t)  =f(\Delta
t)/\int_0^{T_{\rm{o}}} f(\Delta t') d \Delta t'$. 

Fig. \ref{fig:2}b shows that the one-level reversible cascade extracts
less information from the input signal than a network that averages
the input uniformly over time.  Only when $k_{\rm b}$ goes to zero,
and $f(\Delta t) \propto \exp(- k_{\rm b} \Delta t) \approx 1$, does
the network, which now becomes an {\em irreversible} one-level
cascade, implement uniform time averaging and does it extract the same
amount of information.  Intuitively, degrading $X$
destroys information.  While degradation is required
to make a signaling network responsive to new environments, this
example shows that it may be useful to make degradation as weak as possible or to physically separate the receptors and deactivating enzymes (\emph{e.g.} in different
domains on the membrane \cite{zhang2010}), such that $X$ is deactivated only after the response has
been made. 

Signaling networks typically consist of more than one layer, which
makes it possible to sculpt the weighting function $f(\Delta t)$.  As
an illustration, we first consider an \emph{irreversible} cascade
consisting of $N$ layers/species: $dI_i/dt = k_{fi} I_{i-1}$,
where $i=1,\dots, N$ and $I_0 = S$.  Assuming $X(0) \approx 0$,
$X(T_{\rm{o}}) = \int_0^{T_{\rm{o}}} f(T_{\rm{o}}-t) S(t) dt$, where the weighting function now
behaves as $f(\Delta t) \propto \Delta t^{N-1}$.  Such cascades place
more weight on early values of the input signal, which have had more
time to propagate through the network (Fig. \ref{fig:2}c). As a
result, they underutilize (down-weight) the most recent
information in the signal, and indeed, these cascades 
perform worse than a strict average of
the input (Fig. \ref{fig:2}d). 

The above formalism can be generalized to arbitrarily large linear
signaling networks. Multilevel \emph{reversible} cascades have
weighting functions that peak some finite time in the past, balancing
the down-weighting of the signal from the distant past due to the
reverse reactions, with the down-weighting of the signal from the
recent past resulting from the multi-level character of the
network (see supplement). More generally, linear combinations of the weighting
functions for reversible and irreversible cascades can be achieved with multiple cascades that are activated by
the input in parallel and which independently activate the same
effector molecule, as we demonstrate below. Clearly, signaling
networks allow for very diverse weighting functions.

This raises the question whether there exists an optimal weighting
function $f^*(\Delta t)$ that minimizes the error in the estimate of
the ligand concentration.  To this end, we differentiate
Eq. \ref{eq:sigma_c_2} with respect to $\bar{f}$ using Lagrange multipliers
that constrain the integral of $\bar{f}$ to 1, to find the optimal (normalized) weighting
function:
\begin{equation}
\label{eq:f_opt}
\bar{f}^*(\Delta t) = (1-w^*) \dfrac{1}{T_{\rm{o}}} + w^* \dfrac{\delta(\Delta t) + \delta(\Delta t - T_{\rm{o}})}{2}. 
\end{equation}
The first term places equal weight on all prior values of the input,
as assumed in previous studies
\cite{berg1977,bialek2005,levinepre2007,
  levineprl2008,levineprl2010}. The second term, however, places
greater weight on the first and last observed values of the signal,
which are the two signal values that are the least correlated. Indeed,
this is the central result of this manuscript: the optimal weighting
function does \emph{not} correspond to uniform weighting of all signal
values. How much weight is placed on the first and
last points is determined by $w^*=\frac{1}{T_{\rm{o}}/(2 \tau_{\rm c})+1}$,
which decreases from one to zero as the response time $T_{\rm{o}}$ over the
correlation time $\tau_{\rm c}$ increases. 

The optimal weighting function can be
implemented using common network motifs.  For example, the commonly
observed feed-forward loop \cite{shen-orr02} in Fig. \ref{fig:2}g
contains two branches which independently activate $X$.  The left branch, a one-level \emph{reversible} cascade, amplifies
later values of the signal ($t\to T_{\rm{o}}$);  the
right branch, a multilevel \emph{irreversible} cascade,
 amplifies earlier ($t\to 0$) values of the signal.  Together, they produce a
weighting function which selectively amplifies less correlated values
of the input (Fig. \ref{fig:2} g, h), outperforming the uniform
average that could be obtained by reading out node $I_2$ directly.  This simple network illustrates how a spectrum of protein lifetimes and cascade levels can be used to shape weighting functions.

The optimal weighting function $f^*$ also provides the fundamental
limit on the ability of linear signaling networks to measure chemical
concentrations:
\begin{equation}
\label{eq:FundLim}
\sigma_{\hat{c}}^2 [X^*(T_{\rm{o}})] = \frac{\sigma_{\hat{c}}^2 [S] }{T_{\rm{o}}/(2\tau_{\rm c})+1},
\end{equation}
which is obtained by combining Eqs. \ref{eq:sigma_c_2} and
\ref{eq:f_opt}. Eq. \ref{eq:FundLim}
has a simple interpretation: a time series of length $T_{\rm{o}}$ contains an
independent observation every time period of the order of the
correlation time, plus one corresponding to the observation at $t=0$.
Eq. \ref{eq:FundLim} is then the formula for the variance of the mean
of $N=T_{\rm{o}}/(2 \tau_{\rm c}) + 1$ independent, identically distributed random
variables. 

The improvement of the optimal weighting function over uniform
weighting (Eq. \ref{eq:BP}) is maximal when the observation time is
about three correlation times. The maximum improvement over the sample
average is 12\% (Fig. \ref{fig:2}f). While this improvement over the
Berg-Purcell estimate is modest, and smaller than the 50\% improvement that could in principle be obtained by maximum likelihood \cite{wingreen2009,mora2010}, it does show, for the first time, that
simple signaling networks can indeed reach the Berg-Purcell limit, and even exceed it. 

Equally important, our analysis provides a clear perspective on
the integration time.  Clearly, $T_{\rm o}$, the time on which the cell must respond, provides an upper bound on the
integration time. Yet, the processing network weights the input signal
by $f(T_{\rm o}-t)$, which may become zero for $t<T_{\rm o}$. In this
case, the {\em effective} integration time $T_{\rm eff}$ is limited by
the range over which $f(T_{\rm o}-t)$ is non-zero. For example, the
weighting function of the one-level reversible cascade becomes zero on the
time scale $k_{\rm b}^{-1}=\tau_X$, the lifetime of the output
component. This can be (much) smaller than $T_{\rm o}$, in which case
$T_{\rm eff}$ is limited by $\tau_X$: $T_{\rm eff}\sim\tau_X<T_{\rm
  o}$. Essentially, degradation of the output erases memory of the input. However, our study of multi-level reversible cascades shows that in general the
range over which $f(\Delta t)$ is non-zero can be longer than the lifetime of the
individual components. Additional intermediate layers not only change
the form of $f(\Delta t)$, but also extend the range over which it is
non-zero, increasing the integration time over which the output
remembers past signals (see supplement).

Values for the correlation time $\tau_c$ of the input signal and the observation time $T_o$ vary widely across biological systems.  Ligand-receptor half-lives, a key determinant of $\tau_c$, vary at least over more than an order of magnitude, i.e. from milliseconds to an hour \cite{lauffbook,endres2010}.  The cell-cycle time provides an upper bound on $T_o$ \cite{elowitz2005} (e.g. 45 minutes in \emph{E. coli} \cite{elowitz2005} and 100 minutes in yeast \cite{cross2007}), but signaling modules and transcriptional responses can make decisions sooner.  Indeed, $T_o$ is not always significantly larger than $\tau_c$, so that the regime in which linear networks can beat the Berg-Purcell estimate is biologically relevant.  For example, both the MAPK response to EGF stimulation \cite{muller2002,avraham2011} and the NF-$\kappa$B response to TNF stimulation \cite{levchenko2011} peak on the time scale of ligand-receptor debinding (10 minutes \cite{lauffbook} and 30 minutes \cite{grell1998}, respectively).  Additionally, correlation times for gene expression are of the order of the cell cycle time in both \emph{E. coli} and human cells \cite{elowitz2005,alon2006}, suggesting the finite $T_o$ limit is also important for scenarios in which intracellular proteins act as receptors for intracellular signals \cite{bialek2005}.

Interestingly, when $T_o \lesssim \tau_c$, the equilibration time of the signal must be taken into account, since the equilibration time is, according to the fluctuation-dissipation theorem, given by the correlation time, at least when the change in $c$ is small.  Therefore, we end by studying how signaling networks can extract information from non-stationary signals.  We study an input signal generated by $\varnothing
\leftrightharpoons S$ with $S(0) = 0$ and forward and reverse rates
$k_{\rm p} c$ and $k_{\rm r} S$, respectively.  This signal
increases to its steady state value on a time scale $\tau=1/k_{\rm
  r}$, which also equals the steady-state correlation time
$\tau_{\rm c}$.  Extending the procedure in Eqs. \ref{eq:sigma_c_2}
and \ref{eq:f_opt}, the minimal estimation error with a linear
signaling network is $\sigma_{\hat{c}}^2 [X^*(T_{\rm{o}})]
=\frac{\sigma_{\hat{c}}^2 [S]} { T_{\rm{o}}/(2 \tau_{\rm c}) + \ln \left (2
    -e^{-T_{\rm{o}}/\tau_{\rm c}}\right)/2}$ (see supplement). This
shows that less information can be extracted from non-stationary
signals than from stationary ones. To avoid the detrimental effect
of correlations, the optimal weighting function places more weight on
the initial and final points, as for stationary signals. However,
because there is no information at $t=0$, the amplification of early
time points is spread over time points $t<\tau_{\rm c}$ (Fig. S2).
Additionally, the relative amplification of the last time point
increases with decreasing $T_{\rm{o}}$.  Indeed, when $T_{\rm{o}} \ll
\tau_{\rm c}$, no previous signal values are sufficiently uncorrelated
with the most recent one, and 
almost all weight is placed on the final time point $S(T_{\rm{o}})$.

We have studied the ability of linear
signaling networks to extract information from noisy input
signals. While the data processing inequality suggests that it is
advantageous to limit the number of nodes in a signaling network to
minimize the effect of intrinsic noise \cite{deronde2010}, here we
show that there can be a competing effect, in terms of information
processing, in favor of increasing the number of nodes:
better removal of extrinsic noise. Additional nodes make it possible
to sculpt the weighting function for averaging the incoming signal,
allowing signaling networks to reach and even exceed the Berg-Purcell
limit.  Our predictions could be tested experimentally in a controlled setting by using \emph{in vitro} or \emph{in vivo} synthetic signaling networks \cite{lim2010}.  Dual reporter constructs can be used to isolate the effects of extrinsic noise, studied in this Letter, from noise intrinsic to the signaling machinery itself \cite{elowitz2002, bowsher2012}.

This work is part of the research program of the
``\nohyphens{Stichting voor Fundamenteel Onderzoek der Materie}
(FOM)", which is financially supported by the
``\nohyphens{Nederlandse} Organisatie voor Wetenschappelijk Onderzoek
(NWO)." We thank Andrew Mugler and Wiet de Ronde for a careful
reading of the manuscript.

\section{Supplementary Material}

\subsection{Estimation error from linear or linearized signaling networks}

We provide further details into the derivation of Eq. 3 in the main text for linearized signaling networks.  The proof for linear networks, sketched in the main text, follows as a special case.  For linearized signaling networks, $\delta X(T_{\rm{o}}) = \int_0^{T_{\rm{o}}} f_c(T_{\rm{o}}-t) \delta S(t) dt$, where $\delta S(t) = S(t) - E[S(t)|c]$, $\delta X(t) = X(t) - E[X(t)|c]$, and $f$ can depend on $c$ (because the linearization depends on the trajectory the network is linearized about.)  We use $E[Y]$ to denote the expectation of the random variable $Y$.  The dependence of $f$ on $c$ makes the proof of Eq. 3 for linearized networks more subtle than for linear networks.  We start from Eq. 2 in the main text:
\begin{equation}
\label{eq:sig}
\sigma_{\hat{c}}^2 [X(T_{\rm{o}})] =  \sigma_{X(T_{\rm{o}})}^2 / \left(dE[X(T_{\rm{o}})|c]/dc \right)^2
\end{equation}
The variance in the numerator of Eq. \ref{eq:sig} is:
\begin{align}
\label{eq:numer}
 &\sigma_{X(T_{\rm{o}})}^2 = E[ (\delta X(T_{\rm{o}}) )^2 ] \nonumber \\ &=  E\left[ \left( \int_0^{T_{\rm{o}}} f_c(T_{\rm{o}}-t) \delta S(t) dt \right) ^2 \right] \nonumber \\ &= \int_0^{T_{\rm{o}}} \int_0^{T_{\rm{o}}} f_c(T_{\rm{o}}-t_1) E[\delta S(t_1) \delta S(t_2)] f_c(T_{\rm{o}}-t_2) dt_1 dt_2 \nonumber \\ &= \int_0^{T_{\rm{o}}} \int_0^{T_{\rm{o}}} f_c(T_{\rm{o}}-t_1) C(t_1,t_2) f_c(T_{\rm{o}}-t_2) dt_1 dt_2
\end{align}
To determine the denominator of Eq. \ref{eq:sig}, note that $X(T_{\rm{o}}) = E[X(T_{\rm{o}})|c] + \int_0^{T_{\rm{o}}} f_c(T_{\rm{o}}-t) \left( S\left(t\right) - E\left[S\left(t\right)|c\right] \right) dt$.  Taking the expectation at a concentration $c + dc$ yields: $E[X(T_{\rm{o}})|c+dc] = E[X(T_{\rm{o}})|c] + \int_0^{T_{\rm{o}}} f_c(T_{\rm{o}}-t) ( E[S(t)|c+dc] - E[S(t)|c] ) dt$.  Then, because $S$ is stationary:
\begin{equation}
\label{eq:denom}
\dfrac{dE[X(T_{\rm{o}})|c ]}{dc}  = \dfrac{dE[S|c]}{dc} \int_0^{T_{\rm{o}}} f_c(T_{\rm{o}}-t) dt
\end{equation}
Inserting Eqs. \ref{eq:numer} and \ref{eq:denom} into the numerator and denominator, respectively, of Eq. 
\ref{eq:sig}, we find Eq. 3 in the main text:
\begin{align}
\label{eq:result}
&\sigma_{\hat{c}}^2 [X(T_{\rm{o}})]  =  \sigma_{\hat{c}}^2 [ S ] \nonumber \\ & \times \int_0^{T_{\rm{o}}} \int_0^{T_{\rm{o}}} \bar{f}(T_{\rm{o}}-t_1) \bar{C}(t_1,t_2) \bar{f}(T_{\rm{o}}-t_2) dt_1 dt_2.
\end{align}
The integrals of the weighting function in the denominator of Eq. \ref{eq:sig} normalize the weighting functions in the numerator of Eq. \ref{eq:sig}; the
correlation function is normalized by pulling the stationary variance
of the signal $S$ into the prefactor.

\subsection{Multilevel reversible cascades}

\begin{figure}[b]
\includegraphics[width=8.5cm]{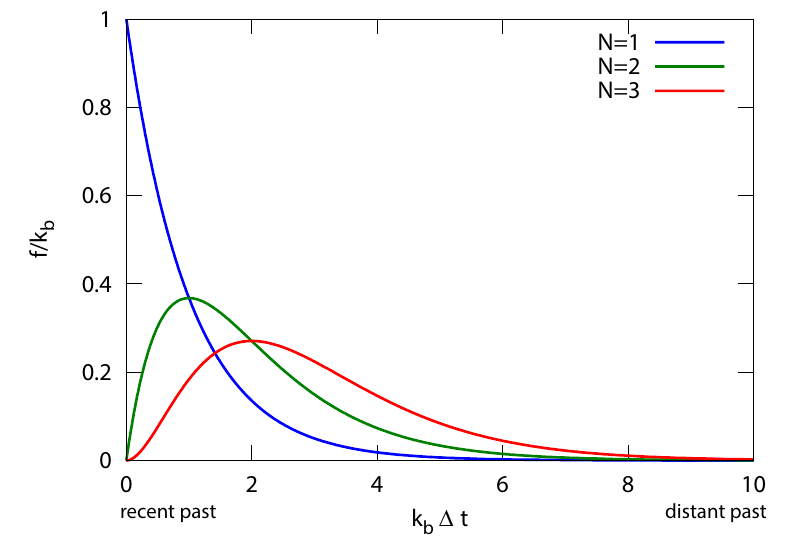}
\caption{ Multilevel reversible cascades extend the integration time over which the output remembers past signals.  A cascade with $N$ levels places maximum weight on past time points $k_b \Delta t^* = n-1 = 0, 1, 2$ for cascades with 1 (blue), 2 (green), or 3 (red) levels, respectively. The plotted weighting functions have been normalized to integrate to 1.}
\label{fig:multireverse}
\end{figure}

We consider a reversible cascade
consisting of $N$ layers/species that are each degraded at the same rate, $k_{\rm b}$: $dI_{\rm i}/dt = k_{\rm fi} I_{i-1} - k_{\rm b} I_i$, where $i=1,\dots, N$, $I_{\rm 0} = S$, and $X = I_{\rm N}$.  Assuming $I_{\rm i}(0) \approx 0$,
$X(T_{\rm{o}}) = \int_0^{T_{\rm{o}}} f(T_{\rm{o}}-t) S(t) dt$, as for the examples in the main text.  For $N=1$, the network is the one-level reversible cascade studied in the main text, with $f(\Delta t) \propto \exp( -k_{\rm b} \Delta t)$.  The one-level reversible cascade places the most weight on the most recent value of the signal ($\Delta t^*=0$).    The weighting function for general $N$, which can be determined by Laplace transforming the governing differential equations, behaves as $f(\Delta t) = \frac{\prod_i k_{\rm fi} }{(n-1)!} \Delta t^{n-1} \exp(-k_{\rm b} \Delta t)$ (Fig. \ref{fig:multireverse}).  The exponential factor, which reflects the reversibility of the cascade, emphasizes the most recent values of the signal; the polynomial factor, which reflects the number of levels, emphasizes older values of the signal.  In combination, the two factors lead to nonmonotonic weighting functions that peak some finite time in the past, $\Delta t^*=\frac{n-1}{k_{\rm b}}$  (Fig. \ref{fig:multireverse}).  As a result, additional levels in a cascade can increase the effective integration time over which the weighting function is nonzero.

By remembering farther into the past, multilevel reversible cascades can improve the performance of signaling networks, provided $T_o$ is not limiting.  Using Eq. 3 in the main text, the estimation error for a one-level reversible cascade, $N=1$, is:
\begin{equation}
\label{eq:single}
\sigma_{\hat{c}}^2 [X(T_{\rm{o}})] =  \frac{\sigma_{\hat{c}}^2 [S]}{\frac{1}{k_{\rm b} \tau_{\rm c}} +1}
\end{equation}
in the limit $T_{\rm o} \gg 1/ k_b$.
This equation is similar to the extrinsic component of the noise-addition rule, after rearrangement of terms \cite{paulsson2003,sorin2006}.  The solution for general $N$ can be obtained in terms of the hypergeometric function but is difficult to analyze.  For a two-level ($N=2$) cascade the estimation error is:
\begin{equation}
\label{eq:twolevel}
\sigma_{\hat{c}}^2 [X(T_{\rm{o}})] =  \frac{\frac{1}{2 k_{\rm b} \tau_{\rm c}}+1}{\left(\frac{1}{k_{\rm b} \tau_{\rm c}} +1\right)^2} \sigma_{\hat{c}}^2 [S]
\end{equation}
which is smaller than the error for the one-level cascade for all finite values of $k_{\rm b} \tau_{\rm c}$.  The improvement is greatest in the limit of slow-decaying molecules, $k_{\rm b} \tau_{\rm c} \to 0$; then, the error of a two-level cascade is half that of a one-level cascade.  More generally, by combining different multilevel cascades that peak at different times in the past, networks can both shape weighting functions and extend the range over which they are nonzero, even when the lifetimes of signaling molecules are limited.  

\subsection{Non-stationary signals}

We consider the non-stationary signal $S(t)$ introduced in the main
text, generated by $\varnothing \leftrightharpoons S$ with $d\mu_S/dt
= k_{\rm p} c - k_{\rm r} \mu_S$ and $S(0)=0$.  The correlation time
is $\tau_{\rm c}= 1/k_{\rm r}$.  Because the signal is nonstationary,
the weighting function $f(t;T_{\rm{o}})$ no longer reduces to $f(T_{\rm{o}}-t)$.  In
what follows we write $f(t)$, where the argument is the time directly
(i.e. the time since the change in environment) and not $\Delta t$, as
for the stationary input signal.  The response of a signaling network
with weighting function $f(t)$ is $X(T_{\rm{o}}) = \int_0^{T_{\rm{o}}} f(t) S(t) dt$.  As
for a stationary signal, the variance of an estimate based on $X(T_{\rm{o}})$
is given by Eq. \ref{eq:sig}, when $X(T_{\rm{o}})$ is linear over the
fluctuations in $S(t)$.  Note that $\mu_{S(t)} = k_{\rm p} c \tau_{\rm c} (1-\exp( -
t/\tau_{\rm c}) )$ so that $dE[X(T_{\rm{o}})|c]/dc$ in the denominator of Eq. \ref{eq:sig} is:
\begin{eqnarray}
\label{eq:denom2}
\dfrac{d\mu_{X(T_{\rm{o}})}}{dc} &=& \int_0^{T_{\rm{o}}}  f(t) \dfrac{d\mu_{S(t)}}{dc} dt
\nonumber \\ &=& k_{\rm p} \tau_{\rm c} \int_0^{T_{\rm{o}}}  f(t) (1-\exp( - t/\tau_{\rm c})) dt
\end{eqnarray}
The numerator of Eq. \ref{eq:sig} is:
\begin{eqnarray}
\label{eq:numer2}
\sigma_{X(T_{\rm{o}})}^2 &=& E\left[ \left( X(T_{\rm{o}}) - \mu_{X(T_{\rm{o}})}\right)^2\right] \nonumber \\ &=&  E\left[ \left( \int_0^{T_{\rm{o}}} f(t) \left(S(t) - \mu_{S(t)} \right)  \right)^2\right] \nonumber \\  &=& \int_0^{T_{\rm{o}}} \int_0^{T_{\rm{o}}} f(t_1) C(t_1,t_2) f(t_2) dt_1 dt_2
\end{eqnarray}
as for a stationary signal, except that the correlation function for the non-stationary signal is:
\begin{equation}
C(t_1,t_2) = k_{\rm p} c \tau_{\rm c} \left( e^{-| t_2 - t_1 | / \tau_{\rm
      c}}-e^{-\max(t_1,t_2) / \tau_{\rm c}}  \right).
\end{equation}
For this non-stationary case, we define a normalized weighting
function $\bar{f}$, which differs from that for the stationary case
presented in the main text:
\begin{equation}
\label{eq:normal}
\bar{f}(t) = \frac{f(t)}{\int_0^{T_{\rm{o}}} f(t') (1-\exp( - t'/\tau_{\rm c})) dt'}
\end{equation}
so that $\int_0^{T_{\rm{o}}} \bar{f}(t) (1-\exp( - t/\tau_{\rm c})) dt = 1$.
Additionally, we define $\bar{C}$ to be the correlation function normalized by the
signal's variance $C(t_1=t_2)$ in steady state, $k_{\rm p} c \tau_{\rm c}$.

Combining Eqs. \ref{eq:sig}, \ref{eq:denom2}, \ref{eq:numer2}, and
\ref{eq:normal}, the analogue of Eq. 3 in the main text for this
non-stationary signal is:
\begin{equation}
\label{eq:nonerr}
\sigma_{\hat{c}}^2 [X(T_{\rm{o}})] = \frac{c}{k_{\rm p} \tau_{\rm c}}  \int_0^{T_{\rm{o}}} \int_0^{T_{\rm{o}}} \bar{f}(t_1) \bar{C}(t_1,t_2) \bar{f}(t_2) dt_1 dt_2
\end{equation}
The prefactor can be interpreted as the error of an estimate of $c$ based only on an instantaneous observation of the signal in steady state (i.e. $T_{\rm{o}} \gg \tau_{\rm c}$): $\sigma_{\hat{c}}^2 [S] =  \sigma_{S}^2 / \left(d\mu_S/dc \right)^2 = k_{\rm p} c \tau_{\rm c} / (k_{\rm p} \tau_{\rm c})^2 = c / (k_{\rm p} \tau_{\rm c})$.

To minimize Eq. \ref{eq:nonerr}, we differentiate with respect to $\bar{f}$, using a Lagrange multiplier to enforce the normalization constraint:
\begin{equation}
\label{eq:inteq}
\int_0^{T_{\rm{o}}} \bar{C}(t_1,t_2) \bar{f}(t_2) dt_2 - \lambda (1-\exp( -
t_1/\tau_{\rm c}))  = 0.
\end{equation}

\begin{figure}[b]
\includegraphics[width=8.5cm]{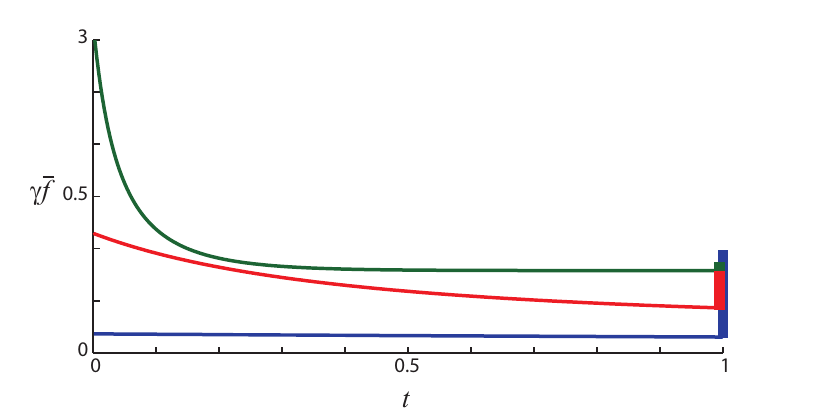}
\caption{ 
Optimal weighting functions for a non-stationary signal.
  The optimal weighting function is plotted for $T_{\rm{o}}=1$ for a signal
  with $\tau_{\rm c}$ = 0.1 (green), 1 (red), and 10 (blue).  For comparison,
  the weighting functions have been renormalized to integrate to 1;
  i.e. we have multiplied by a factor $\gamma$ so that $ \gamma
  \int_0^{T_{\rm{o}}} \bar{f} dt = 1$.  The delta functions at time $T_{\rm{o}}$ are
  truncated for illustration, with height equal to their respective
  coefficients.  Some minimal weight is placed on all points; how much
  depends on $\tau_{\rm c}$ and $T_{\rm{o}}$.  Then additional weight is placed on
  early and late data points, because of correlations in the input
  signal. The final time point dominates the estimate when $T_{\rm{o}} \ll
  \tau_{\rm c}$ (blue curve).  }
\label{fig:model}
\end{figure}

One way to solve Eq. \ref{eq:inteq} is to differentiate three times
with respect to $t_1$, resulting in an ordinary differential equation
after substitution of intermediate derived equalities (see
ref. \cite{integralbook} for a discussion of this method for solving
integral equations).  The solution is (Fig. \ref{fig:model}):
\begin{equation}
\label{eq:nonopt}
\bar{f}^*(t) = c_1  \frac{  1 } {\left( 2 - e^{-t/\tau_{\rm c}}  \right)^2 } + c_2 \delta(t-T_{\rm{o}}) 
\end{equation}
with 
\begin{eqnarray}
c_1 &=& \frac{4} {T_{\rm{o}} +\tau_{\rm c} \ln\left(2 -e^{-T_{\rm{o}}/\tau_{\rm c}}\right) } \\
c_2 &=& c_1 \frac{\tau_{\rm c} }{ 2 \left( 2  - e^{-T_{\rm{o}}/\tau_{\rm c}} \right)}
\end{eqnarray}
Note that the weighting function has units of 1/time, $c_1$ has units
of 1/time, $c_2$ has no units, and the delta function has units of
1/time.  The weight placed on the final time point grows relative to
the weight placed on other points as $T_{\rm{o}}/\tau_{\rm c}$ decreases, as measured
by its contribution to the integral of the weighting function.  The
first term approaches a constant weight for $t>\tau_{\rm c}$.

The corresponding minimal estimation error is, as in the main text:
\begin{equation}
\label{eq:nonperf}
\sigma_{\hat{c}}^2 [X^*(T_{\rm{o}})] = \frac{\sigma_{\hat{c}}^2 [S]} { T_{\rm{o}}/(2
  \tau_{\rm c}) +  \ln \left (2  -e^{-T_{\rm{o}}/\tau_{\rm c}}\right)/2}
\end{equation}
The short and long time limits of Eq. \ref{eq:nonperf} can be
motivated with simple arguments.  For short observation times $T_{\rm{o}} \ll
\tau_{\rm c}$, the estimate is essentially constructed from $S(T_{\rm{o}})$
only, since Eq. \ref{eq:nonopt} indicates that all weight is placed on
the final time point in that limit.  Because no $S$ molecules decay on
the short time $T_{\rm{o}} \ll \tau_{\rm c}$, the number of $S$ molecules at time $T_{\rm{o}}$
is Poisson distributed with arrival rate $k_{\rm p} c$, mean $k_{\rm p} c T_{\rm{o}}$, and
variance $k_{\rm p} c T_{\rm{o}}$.  The variance of an estimate based on $S(T_{\rm{o}})$ is
then, from Eq. \ref{eq:sig}, $(k_{\rm p} c T_{\rm{o}})/(k_{\rm p} T_{\rm{o}})^2 =
c/(k_{\rm p} T_{\rm{o}})$, the
short time approximation of Eq. \ref{eq:nonperf}.  The long time
approximation of Eq. \ref{eq:nonperf} is $\sigma_{\hat{c}}^2 [S] /
(T_{\rm{o}}/(2 \tau_{\rm c}))$, identical to the long-time approximation for an
estimate based on a stationary signal of equivalent duration (see
Eq. 5 in the main text).  The effect of the non-stationarity is washed
out on long time scales.  For finite times, the non-stationary signal
contains less extractable information than a stationary signal of
equal duration.

\end{document}